\documentclass[a4paper,11pt]{article}
\pdfoutput=1 

\usepackage{jheppub} 

\usepackage[T1]{fontenc} 

\newcommand{\be}{\begin{equation}}
\newcommand{\ee}{\end{equation}}

\title{\boldmath  Exact Casimir interaction  of perfectly conducting three-spheres in four euclidean dimensions.}

\author{Giuseppe Bimonte}

\affiliation{ Dipartimento di Fisica E. Pancini, Universit\`{a} di
Napoli Federico II, Complesso Universitario
di Monte S. Angelo,  Via Cintia, I-80126 Napoli, Italy}
\affiliation{INFN Sezione di Napoli, I-80126 Napoli, Italy}

\emailAdd{bimonte@na.infn.it}

\abstract{Exploiting conformal symmetry, we derive  a simple exact formula for the classical  electromagnetic Casimir interaction  of two perfectly conducting three-spheres, including the sphere-plate geometry as a special case,
in four euclidean dimensions.  
We verify that  the short distance expansion of the Casimir energy agrees to leading order with the Proximity Force Approximation (PFA), while the next-to-leading-order is in agreement with a recently proposed derivative expansion of the Casimir energy. At the next-to-next-to-leading order we find a non-analytic correction to PFA, which for a sphere-plate system is of the order of $(d/R)^{3/2} \log(d/R)$, where $d$ is the separation and $R$ the sphere radius. }

\begin{document} 
\maketitle
\flushbottom

\section{Introduction}
\label{sec:intro}

The Casimir effect \cite{Casimir48} is the tiny force between two neutral macroscopic polarizable  bodies, that originates from quantum and thermal fluctuations of the electromagnetic (em) field in the region of space bounded by  the surfaces of the two bodies.  This is one of the rare manifestations of quantum physics at the macroscopic scale, like superconductivity and superfluidity. The last two decades witnessed a strong resurgence of interest in the Casimir effect,  spurred by a new wave of experiments  which measured the Casimir force  with unprecedented precision. For reviews, see \cite{book1,parse,book2}. 

The (em) Casimir effect represents just an example of more general fluctuation-induced forces \cite{kardar} that arise  when two objects are embedded in a correlated  medium.  In recent years much attention has been  attracted by  so-called {\it critical} Casimir forces \cite{degennes,Krech} that  originate from  classical thermal fluctuations of a fluid in the vicinity of a critical point, where correlation lengths are macroscopic.  Very recently, critical  Casimir forces have been observed in helium~\cite{Garcia:1999}
and in binary liquid mixtures~\cite{Law:1999,Pershan:2005,Hertlein:2008}.(In the rest of this paper, by Casimir effect we shall denote just the original em effect discovered by Casimir).

A distinctive feature of Casimir forces is their non-additivity. As a result of this feature Casimir forces depend in a complicated way on the geometry and material properties of the intervening objects,  and because of that they are very hard to compute in non-planar geometries.  In his pioneering paper Casimir   studied the highly idealized system of two perfectly conducting  large parallel plates in vacuum  at zero temperature,  for which he obtained an attractive force of magnitude:
\be
F_C=\frac{\pi^2 \hbar c}{240\,d^4}\,A\;,
\ee 
with $A$ the area of the plates, and $d$ the separation.
The theory of the Casimir effect for real material surfaces was developed a few years later by Lifshitz \cite{lifs}, who derived a formula for the   force between two plane-parallel  dielectric slabs at finite temperature.  Unfortunately the planar geometry studied by Casimir and Lifshitz is extremely hard to implement, due to the insurmountable difficulty of keeping parallel two macroscopic plates posed at a submicron distance from each other.  To avoid these problems, practically all  present  Casimir experiments (with the notable exception of \cite{gianni,antonini} where the plane-parallel geometry was used) adopt the sphere-plate geometry for which the parallelism issue does not arise.  Until recently, nobody knew how to compute the Casimir interaction between two non-planar surfaces, like a sphere and a plate. The  commonly used approximation to deal with curved surfaces  was the old-fashioned Proximity-Force-Approximation (PFA) introduced long ago by Derjaguin \cite{Derjaguin},  which consists in   averaging  the force between two parallel plates as provided by Lifshitz formula, over the (appropriately defined) local surface-surface separation. The PFA is believed to provide the leading term of the small-distance expansion of the Casimir energy between two smooth surfaces in the limit of vanishing separation. An important breaktrough came about ten years ago, when a scattering formula was found that in principle allows to  compute Casimir forces between dielectric objects of any shape \cite{sca1,sca2}. While some of these results  had been  discovered much earlier by Langbein \cite{langbein}, but soon came into oblivion, concrete analytical and numerical results have been obtained only recently. Indeed, the experimentally most relevant sphere-plate system was only treated very recently \cite{emig,lambr}.  

Despite the tremendous theoretical advancement represented by the scattering formalism, approximate methods like the PFA still retain a great practical importance. This is so because the exact scattering formula for the Casimir energy is   viable only for relatively large sphere-plate separations $d$, but  becomes untractable even numerically in typical experimental situations where $d/R \sim 10^{-3} \div 10^{-4}$, with $R$ the sphere radius. As a matter of fact, the PFA is still widely used today to interpret theoretically  current precise Casimir experiments. The  resolution of much debated issues in Casimir physics, like the  magnitude of the thermal contribution to the Casimir force \cite{book2,brevik,bimonteprec,bimontehide,bimontehide2}, depends crucially on our ability to obtain reliable predictions of the Casimir force between metallic conductors.  In order to assess the theoretical error introduced by the PFA, several researchers have endeavored to compute the next-to-leading-order (NTLO) term in the small-distance expansion of the Casimir energy, i.e. the  first correction beyond PFA. There are presently two approaches to achieve this goal. The first one is rigorous, but extremely laborious as it involves working out the asymptotic small-distance expansion of the exact scattering formula. By following this route, the   NTLO energy has been computed for the cylinder/plate and the sphere/plate geometries, initially for a free scalar field obeying Dirichlet (D) boundary conditions (bc) \cite{bordag1}, and then for the em field with ideal metallic bc \cite{bordag2}. Later  the same approach was applied to  a free scalar field obeying  D, Neumann (N) and mixed ND bc on two parallel cylinders \cite{teo1}. An alternative and computationally much simpler route to compute  the NTLO energy was introduced in \cite{fosco1}, based on  a derivative expansion (DE)   in the local separation between surfaces for the force between gently curved bodies.  In \cite{fosco1} the DE  was applied to a D scalar field in the cylinder and sphere/plate geometries,  giving results in agreement with those obtained by scattering methods  in \cite{bordag1}. The DE for the more general case of two curved surfaces was later  worked out in \cite{bimonte1} for the em field with perfect-conductor bc, as well as for a scalar field obeying N and mixed DN bc.   Interestingly, the first correction beyond PFA for the perfect-conductor sphere/plate geometry obtained in \cite{bimonte1}  by using the DE   was in disagrement with that reported in \cite{bordag2}: while the DE predicted an analytic  correction $\sim d/R$,   a larger logarithmic $\sim d/R\,\log(d/R)$ correction had been found in \cite{bordag2}.  
A successive recalculation by some of the authors of \cite{bordag2} detected a sign mistake in their orginal computation, and finally  led to full agreement  with the DE expansion also in em  and N cases.  The DE for a D and N scalar at zero and finite temperature in any number of space-time dimensions   was worked out in \cite{fosco2}, while the experimentally important case of dielectric curved surfaces at finite temperature  is presented in \cite{bimonte2}. The DE has been  also used to study curvature effects in
the Casimir-Polder interaction of a particle with a gently curved surface \cite{CPbimonte1,CPbimonte2}. The same method has been used very recently to estimate the shifts of the rotational
levels of a diatomic molecule due to its van der Waals interaction with a curved dielectric surface \cite{CPbimonte3}.

In view of the complicated  shape dependence of the Casimir interaction in non-planar geometries, exact solutions whenever available are much valuable. With their help one may hope on one hand to better understand the behavior of the Casimir interaction in the experimentally important limit of small-distances, and on the other hand they provide useful testing grounds for available approximation schemes like the PFA or the DE.  In the existing literature there exist only  a bunch of  exact solutions. As a matter of fact, all  (but one) exact solutions found so far  are for systems possessing conformal invariance \cite{difrancesco}. The first example is provided by the Casimir force between two spherical particles in a critical fluid \cite{eisen}.  This system can be mapped to two concentric spheres    by means of  a special conformal transformation. By exploiting such a transformation, the authors  of \cite{eisen} obtained a simple formula for the classical Casimir energy
of free scalar fields satisying conformally invariant bc (i.e. D bc and Robin-type bc) on the surfaces of the spheres, in any number of space dimensions.  The   same solution for a D scalar, in the special case of the sphere-plate geometry, has been recently re-derived in \cite{teo2} by performing a similarity transformation in the  scattering formula.  As it is well known, the group of local conformal transformations in two dimensions is infinite dimensional \cite{difrancesco}. The full power of 2D local conformal transformations   has been recently exploited in \cite{bimonte3} to derive an exact formula for  the Casimir interaction between two objects of {\it any} shape embedded in a two-dimensional critical fluid. By using this formula, it has been shown that the sign of the critical Casimir force between two periodically deformed one-dimensional boundaries can be reversed simply by shape deformation \cite{bimonte4}. 

In the experimentally relevant em case, so far the only known exact solution in a non-planar geometry is for two metallic  spheres and a sphere-plate, either grounded or ungrouded, in the high-temperature limit \cite{bimonte5}. In this limit, the quantum Casimir interaction between two   conductors  reduces to the  classical  Casimir interaction  for a scalar field (corresponding to the scalar em potential). The scalar field is subjected to D bc in the grounded case, and  to Drude bc in the ungrounded case. While the former case had been easily solved earlier \cite{eisen} using conformal invariance, the non-conformally invariant   Drude case  is considerably more difficult.  The solution was derived in \cite{bimonte5} by performing a sequence of  complicated similarity transformations on the exact scattering formula. We note that the  classical Casimir interaction between  a sphere and a plate  with Drude bc had been investigated earlier numerically by   evaluating  the scattering formula to very high (several thousandt) multipole order \cite{lambr2}.  The exact formula derived in \cite{bimonte5} is in perfect agreement with the numerical results of \cite{lambr2}, and once expanded  at short distances it displayed an intricate structure of deviations from the commonly employed proximity force approximation. The differences between the high-temperature Casimir interactions of grounded vs ungrounded conductors  are further discussed in \cite{fosco3}. 

In this paper we obtain a simple formula for the exact classical Casimir interaction betwee two perfectly conducting three-spheres,   and between a three-sphere and an hyperplane  in four euclidean dimensions. Our derivation  is based on the well-known conformal invariance  of (vacuum) Maxwell  equations  in four dimensions  \cite{wald}, and exploits conformal invariance of perfect conductor bc. While the solution we find has no direct physical meaning, it  presents a certain interest on its own. Its interest stems from the well established  correspondence between zero-temperature {\it quantum} statistical systems in three spatial dimensions and {\it classical} statistical  systems in four euclidean dimensions \cite{itzy}. According to this correspondence, the {\it quantum}   Casimir interaction of two perfectly conducting (two dimensional) surfaces $\Sigma_1^{(2)}$ and $\Sigma_2^{(2)}$ in physical space $E^{(3)} \equiv \{x_1,x_2,x_3 \}$ at zero temperature,   is the same as the {\it classical}  Casimir interaction of two (three-dimensional) perfectly conducting cylindrical surfaces ${\cal C}^{(3)}_1=R \times \Sigma_1^{(2)}$ and ${\cal C}^{(3)}_2=R \times \Sigma_2^{(2)}$ embedded in four-dimensional euclidean space $E^{(4)} \equiv \{x_1,x_2,x_3,x_4 \}$ (at finite temperature $T$ the correspondence  still holds provided that one imposes periodic bc in the fourth euclidean direction $x_4$, the period $\ell$ being determined by the temperature according to the relation $\ell=\hbar c/(k_B T)$).
Clearly  three-spheres $S^{(3)}$ in $E^{(4)}$  do not correspond to any  surface $\Sigma^{(2)}$ in physical space, and therefore their classical Casimir energy    has no direct physical interpretation in terms of the quantum interaction between two conductors in physical space $E^{(3)}$.   Despite  the unphysical geometry of the boundaries, the study of the Casimir interaction of three-spheres in four euclidean dimensions may provide useful information on the $T=0$ quantum interaction of conductors in physical space.  In particular, in the limit of small separations, it is not unreasonable to imagine that   curvature corrections to PFA  might have a similar structure for both problems.  
In addition to that, the exact solution  provides a valuable opportunity to test quantitatively the accuracy of the approximation schemes described earlier,  i.e. the PFA and the DE.        

The plan of the paper is as follows: in Sec. \ref{sec:scat} we briefly review the scattering formalism for the Casimir effect, and in Sec. \ref{sec:spheres} we use the scattering formula to compute the Casimir energy for the conformal system of two perfectly conducting three-spheres in four euclidean dimensions. In Sec. \ref{sec:smalld} we work out the small distance expansion of the exact Casimir energy for two spheres, derived in Sec. \ref{sec:spheres} and prove that its leading term agrees with the PFA. In Sec. \ref{sec::DExpa} we introduce  the DE and show that it reproduces the NTLO term of the small-dispance expansion of the exact Casimir energy. In Sec. \ref{sec:DENN} we prove that  the DE for the em field breaks down  after the second order, and therefore it cannot be used to compute the  next-to-next-to-leading-order (NNTLO) term in the expansion of the Casimir energy. In Sec. \ref{sec:DENN} we  present arguments showing that the fourth order DE may exist for other field theories, like a D scalar field, while it does not exist for a N scalar. In Sec. \ref{sec:conc} we present our conclusions. Finally, in the Appendix we briefly discuss scattering of em waves by a perfectly conducting sphere in four euclidean dimensions.

\section{The scattering formula for the Casimir free energy}
\label{sec:scat}

According to the scattering approach \cite{emig,lambr} the Casimir interaction energy ${\cal F}$ between two objects can be expressed in terms of the respective scattering amplitudes $\hat{{\cal T}}_1$ and $\hat{{\cal T}}_2$  and translation operators $\hat{{\cal U}}^{12}$ and  $\hat{{\cal U}}^{21}$, that translate the scattering solution from the coordinate system of one object to the one of the other object.  In the  classical limit, the scattering approach  yields the Casimir energy (representing the zeroth-order Matsubara term of the full quantum Casimir energy) 
\be
{\cal F}=\frac{k_B T}{2}\ln \det[1-\hat{{\cal U}}^{21}\, \hat{{\cal T}}_1 \, \hat{{\cal U}}^{12}\,\hat{{\cal T}}_2]\;.\label{casen}
\ee
In principle, the above Equation permits to compute the (classical) Casimir energy for two objects of any shape. In practice, evaluating Eq. (\ref{casen}) is very hard, for two reasons. On one hand,  the scattering operators $\hat{{\cal T}}_1$ and $\hat{{\cal T}}_2$  of the individual objects are unknown, and they are in general very difficult to compute, even numerically, unless the bodies have very simple shapes like spheres or cylinders, for which the scattering problem can be solved analytically in a suitable multipole basis.  On the other hand, even when the scattering operators are known, or can be computed numerically,  it is rarely the case that the scattering and the translation matrices are simultaneously diagonal in some multipole basis. As a rule, they are non-diagonal infinite-dimensional matrices, and therefore Eq. (\ref{casen}) involves in general evaluation of infinite-dimensional determinants, which is of course an impossible task in general. The common practice is to truncate the   determinant to some finite multipole order, and to carefully examine convergence of the result as an increasing number of multipoles is included. The problem is  that the necessary number of multipoles increases very rapidly as the (minimum) distance $d$ of the bodies decreases, in comparison with their characteristic size $L$.  Roughly, convergence is achieved when the number of multipoles becomes of the order of $L/d$, i.e. $10^3 \div 10^4$ in typical experimental situations, which surely represents a non-trivial challenge. 

\section{The Casimir energy for two three-spheres}
\label{sec:spheres}

We  shall use the general Eq. (\ref{casen}) to estimate the Casimir energy for two  (and non-overlapping) perfectly conducting three-spheres $S_1^{(3)}$ and $S_2^{(3)}$   in four euclidean dimensions. In the Appendix, it is shown that in a suitable basis of spherical waves with origin at the center of the three-sphere, the scattering amplitudes for  the internal and external scattering of em waves  by a perfectly conducting three-sphere are both  equal to minus the identity matrix (see Eq. (\ref{scatmat})).  Having determined the scattering amplitude, the next step towards computing the Casimir energy is to determine the matrix elements of the translation operators $\hat{{\cal U}}^{12}$ and $\hat{{\cal U}}^{12}$ in  a suitable  basis of waves   attached to the two spheres. By definition \cite{emig} the matrix  ${\cal U}_{\alpha \alpha'}^{21}$ (${\cal U}_{\alpha'\alpha}^{21}$)   connects the scattered fields $A^{({\rm scat|1})}_{i|\alpha'}$ ($A^{({\rm scat|2})}_{i|\alpha}$) relative to sphere one (two)  to the incoming waves $A^{({\rm in|2})}_{i|\alpha}$ ($A^{({\rm in|1})}_{i|\alpha'}$) relative to sphere two (one):
\be
A^{({\rm scat|1})}_{i|\alpha'}  =\sum_{\alpha}  A^{({\rm in|2})}_{i|\alpha}\, {\cal U}_{\alpha\alpha'}^{21}\;,
\ee 
\be
A^{({\rm scat|2})}_{i|\alpha} =\sum_{\alpha'}  A^{({\rm in|1})}_{i|\alpha'}\, {\cal U}_{\alpha'\alpha}^{12}\;,
\ee  
where $\alpha$ and $\alpha'$ label the basis elements for the two spheres.
Suppose now, to be definite, that the two   spheres $S_1^{(3)}$ and $S_2^{(3)}$ are placed one outside the other. If,  following the normal procedure \cite{emig,lambr},  we used as a wave basis the simple spherical basis described in Eq. (\ref{exter}) attached to the respective centers of the two spheres,  this would immediately result into non-diagonal translation matrices, rendering the computation of the Casimir energy very hard. An alternative route is however possible.  As it is explained below, conformal invariance of the problem  allows us to    express the Casimir energy of two non concentric spheres in terms of the  Casimir energy of a conformally equivalent system of two {\it concentric} spheres. The Casimir energy of the latter highly symmetric system is very easy to compute, thanks to the fact that its translation matrices  are diagonal and easy to compute.

\subsection{Two concentric three-spheres}  

For the highly symmetric configuration of two concentric spheres of radii $R_-$ and $R_+$ (for definitiness we take $R_-   < R_+$), the  basis of spherical waves for the two spheres have the same origin, coinciding with their common center. This feature   enormously simplifies the problem. Because of that, the translation matrices ${\cal U}_{nlmp,n'l'm'p'}^{21}$ and ${\cal U}_{nlmp,n'l'm'p'}^{12}$ are both {\it diagonal} in the basis (\ref{four}), and with our normalization of the respective incoming and scattered waves, it can be easily verified that they have matrix elements: 
\be
{\cal U}_{nlmp,n'l'm'p'}^{21}={\cal U}_{nlmp,n'l'm'p'}^{21}=\left(\frac{R_-}{R_+} \right)^n \;\delta_{nn'} \delta_{l l'} \delta_{mm'} \delta_{pp'}\;.\label{tramat}
\ee
By substituting these translation matrices together with the scattering matrices Eq. (\ref{scatmat}) into Eq. (\ref{casen}),  we obtain the following simple formula for the classical Casimir energy of two perfectly conducting concentric spheres:
\be
{\cal F}=k_B T \, \sum_{n \ge 2} (n^2-1) \log (1-\rho^{2n})\;,\label{EMene}
\ee
where we set $\rho=R_-/R_+ < 1$. 

\subsection{Two non concentric three-spheres} 

Consider two perfectly conducting non-concentric (non-overlapping) three-spheres $S_1^{(3)}$ and $S_2^{(3)}$ of radii $R_1$ and $R_2$. The key observation that allows us to compute their Casimir interaction is that any two such spheres may be obtained by conformally mapping \cite{eisen} the highly symmetric system of two concentric spheres considered in the previous section. 
The special conformal map $\phi: E^{(4)} \rightarrow E^{(4)}$ that achieves this goal is:   
\be
\frac{\bf r'}{r'^2}= \frac{{\bf r}+{\bf R}}{|{\bf r}+{\bf R}|^2}-\frac{\bf R}{2\,R^2}\;.\label{conf}
\ee
where ${\bf R}$ is an arbitrary fixed four-vector. It can be verified that for $R_- < R < R_+$ the concentric three-spheres of radii $R_+$ and $R_-$ in the ${\bf r}$ euclidean space get respectively mapped to
the three-spheres $S_1^{(3)}$ and $S_2^{(3)}$  of radii $R_1$ and $R_2$, placed one {\it outside} the other in the  ${\bf r}'$ euclidean space, whose centers  lie along the straight line of the vector ${\bf R}$. The radii $R_1$ and $R_2$  are
\be
R_{1}=4 R \frac{R R_{+}}{R_+^2-R^2 }\;,\;\;\;\;R_{2}=4 R \frac{R R_{-}}{R^2-R_-^2 }\;,
\ee 
and the minimum distance $d$ between $S_1^{(3)}$ and $S_2^{(3)}$ is
\be
d=4 R \frac{R(R_+-R_-)}{(R+R_+)(R+R_-)}\;.
\ee 
By adjusting the values of $R_+,R_-$ and $R$ it is possible to obtain any value for $R_1$, $R_2$ and $d$.
We note that the sphere-plate system in ${\bf r}'$ space is recovered in the limit $R \rightarrow R_+$, in which case the $R_+$ sphere gets mapped to the three-plane through the origin of ${\bf r}'$ space, perpendicular to ${\bf R}$.  
For $R < R_-$ and for $R> R_+$ the concentric spheres of radii $R_-$ and $R_+$ in the ${\bf r}$   space get mapped to two non-concentric spheres in the ${\bf r}'$ space that are placed one {\it inside} the other. For brevity, we shall only consider below the case of two spheres placed one outside the other in ${\bf r}'$ space, i.e. $R_- < R < R_+$, the extension to the case of two non-concentric spheres one placed inside the other  being straightforward.

At this point, conformal invariance of Maxwell equations in four dimensions  enter into play. Conformal invariance of Maxwell  equations implies that the pull-back by the map $\phi^{-1}$  of the ingoing and scattered fields for two concentric spheres in ${\bf r}$ space (as given by Eqs. (\ref{exter}) and (\ref{inter}) in the Appendix)  constitute a basis of ingoing and scattered em waves in ${\bf r}'$ space. More precisely we set
\be
A^{({\rm in|1})}_{a|nlmp}=(\phi_*^{-1} A)^{({\rm in|int})}_{a|nlmp}\;,\;\;\;\;A^{({\rm scat|1})}_{a|nlmp}=(\phi_*^{-1}  A)^{({\rm scat|int})}_{a|nlmp}\;,\label{basis1}
\ee 
\be
A^{({\rm in|2})}_{a|nlmp}=( \phi_*^{-1}  A)^{({\rm in|ext})}_{a|nlmp}\;,\;\;\;\;A^{({\rm scat|2})}_{a|nlmp}=( \phi_*^{-1}  A)^{({\rm scat|ext})}_{a|nlmp}\;.\label{basis2}
\ee 
Since the perfect conductor bc Eq. (\ref{bc2}) (or the equivalent bc on the potential Eq. (\ref{bc3})) are conformally invariant too (in fact they are invariant under any diffeomorphism), it follows that the scattering matrix Eq. (\ref{scatmat}) is conformally invariant as well. Thus, in the basis (\ref{basis2}) the scattering matrices of the non-concentric spheres $S_1^{(3)}$ and $S_2^{(3)}$ both have the simple diagonal form in Eq. (\ref{scatmat}). Finally, one observes that since the pull-back is a linear transformation, translation matrices are also preserved by the conformal map, and thus the translation matrices  ${\cal U}_{nlmp,n'l'm'p'}^{21}$ and ${\cal U}_{nlmp,n'l'm'p'}^{12}$ for $S_1^{(3)}$ and $S_2^{(3)}$  written in the conformal basis Eqs. (\ref{basis1}) and (\ref{basis2}) are respectively identical to the diagonal translation matrices for two concentric spheres Eq. (\ref{tramat}).   The conclusion of the above considerations is that the Casimir free energy of the spheres $S_1^{(3)}$ and $S_2^{(3)}$ is identical to the Casimir energy of the corresponding conformal system of two concentric spheres, and is thus provided by the simple formula Eq. (\ref{EMene}) in which the parameter $\rho$ is now expressed in terms of the geometric parameters $R_1, R_2$ and $d$ characterizing $S_1^{(3)}$ and $S_2^{(3)}$.  The explicit relation between these parameters is as follows:
\be
\kappa=\frac{1}{2}(\rho+\rho^{-1})\;,\label{kappa1}
\ee 
where
\be
\kappa=\frac{s^2-R_1^2-R_2^2}{2 R_1 R_2}\;,\label{kappa2}
\ee
with $s$ the center-to-center distance $s=d+R_1 +R_2$ between $S_1^{(3)}$ and $S_2^{(3)}$. Thus, Eq. (\ref{EMene}), together with the above two relations provide the complete solution of the Casimir problem for two three-spheres placed one oustide the other. The case of two non-concentric three-spheres placed one inside the other is handled in a completely analogous way, the only difference consisting in the algebraic form of the relations connecting $\rho$ to $R_1$, $R_2$ and $d$, that can be easily worked out starting from Eq. (\ref{conf}).    
 
\section{Small-distance expansion of the Casimir energy}
\label{sec:smalld}
 
\begin{figure}[htbp]
\centerline{\includegraphics [width=.6\columnwidth]{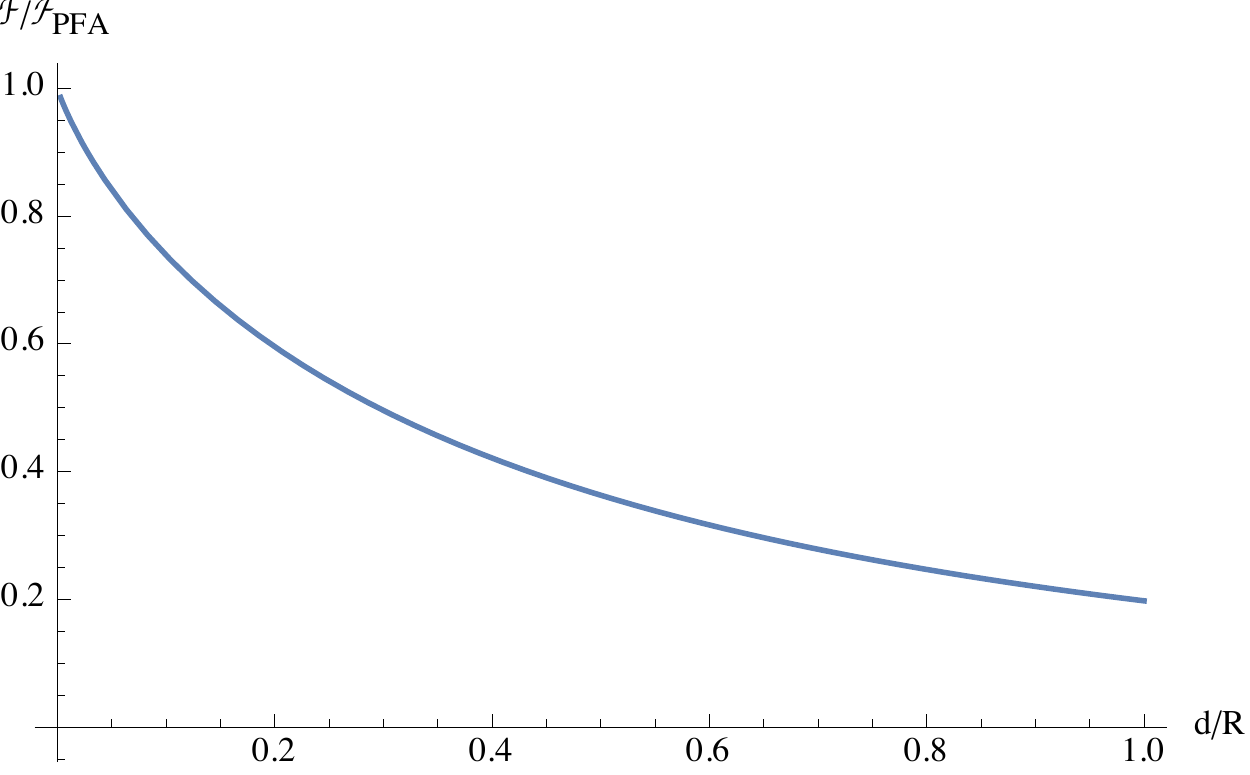}}
\caption{\label{figPFA}  Sphere-plate Casimir energy normalized by the PFA. }
\end{figure}

\begin{figure}[htbp]
\centerline{\includegraphics [width=.6\columnwidth]{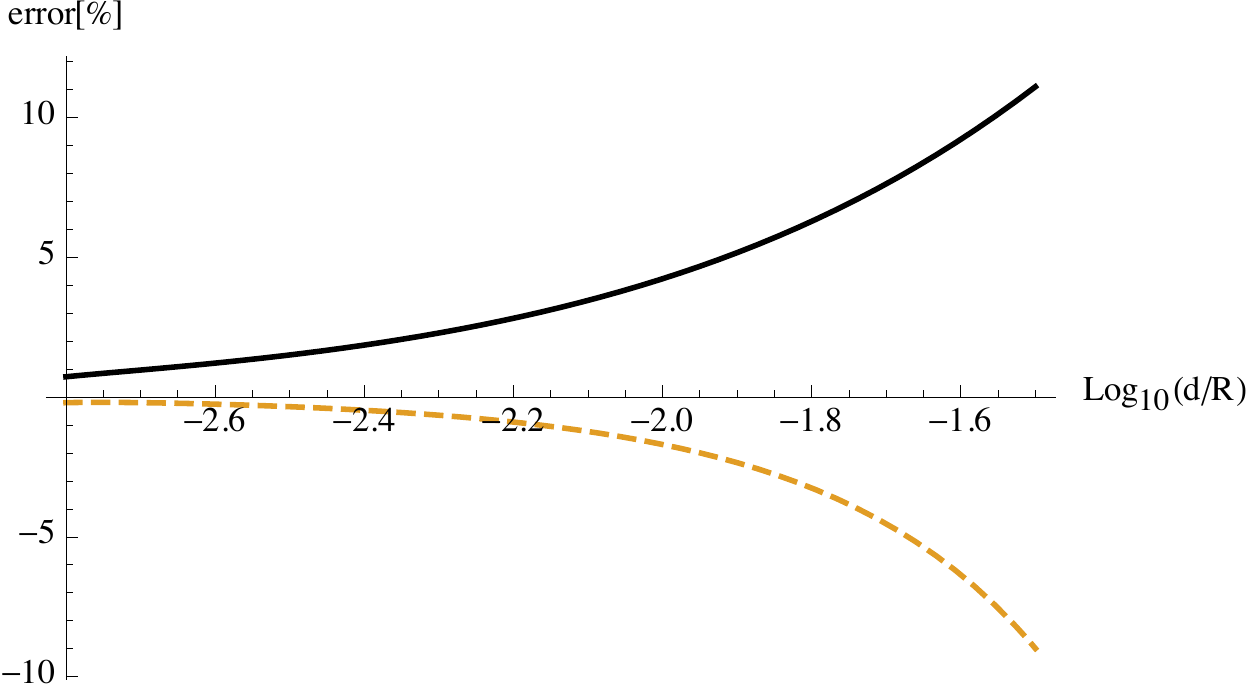}}
\caption{\label{error}   Percent error on the Casimir energy made by using the PFA (solid line) and the DE (dashed line) for the sphere-plate Casimir energy.}
\end{figure}

With an exact expression for the Casimir energy, one can compute explicitly the interaction in the experimentally important limit of short distances. According to Eqs. (\ref{kappa1}-\ref{kappa2}) the small distance limit corresponds to $\rho$ close to one. We thus set $\rho=\exp(-\mu)$ and  compute the series in Eq. (\ref{EMene}) for small $\mu$ using the Abel-Plana formula \cite{book2}. Upto exponentially small terms we obtain the exact expansion:
\be
{\cal F}=-k_B T  \left[ \frac{\pi^4}{360 \mu^3}+\frac{\pi^2}{12 \mu}+\frac{1}{2} \log(\mu/\pi) -\frac{\zeta(3)}{4 \pi^2}+\frac{11}{120} \mu\right]  \;.\label{asen1}
\ee
Consider for simplicity  a three-sphere of radius R at distance $d$ from a three-plane.   Upon substituting $R_2=R$ and taking $R_1 \rightarrow \infty$ into Eqs. (\ref{kappa1}) and (\ref{kappa2}) we obtain
\be
\mu= \log[1+x +\sqrt{x (2+x)}]\;,\label{dsuR}
\ee
where we set $x=d/R$.
Substitution of Eq. (\ref{dsuR}) into Eq. (\ref{asen1}) results in the following small-distance expansion for the  sphere-plate energy
 $$
 {\cal F} = -k_B T \left\{ \,\frac{\sqrt{2}\,\pi^4}{1440\,x^{3/2}} \left[ 1+ \left(\frac{1}{4}-\frac{60}{\pi^2} \right)\, x   
  +\left( \frac{132}{\pi^4}-\frac{7}{480}  - \frac{5}{\pi^2}\right) x^2\right.\right.
$$
\be+\left.\left. \frac{30 \sqrt{2}} {\pi^4} x^{5/2} +\left( \frac{457}{120960} - \frac{11}{\pi^4} + \frac{17}{24 \pi^2}\right) x^3 -\frac{11 }{\sqrt{2} \pi^4} x^{7/2}+o(x^4)\right]+
\frac{1}{4} \log (2 x/\pi^2)+\frac{\zeta(3)}{4 \pi^2}\right\}
 \; \label{EMexp}
\ee
The above small-distance expansion has been derived from the exact Casimir energy Eq. (\ref{EMene}). It is interesting to check if its leading term is correctly reproduced by the PFA. The starting point of the PFA is the (classical) Casimir energy (per unit three-volume) ${\cal F}^{(4)}_{pp}(d)$ for two parallel 3-planes in $E^{(4)}$ at distance $d$.  On dimensional ground, ${\cal F}^{(4)}_{pp}(d)$ must be of the form ${\cal F}^{(4)}_{pp}(d)=k_B T\, u/d^3$, where $u$ is a pure number. We do not need to compute the constant $u$, because its value can be easily inferred 
from the well-known formula for the quantum zero-temperature Casimir energy between two paralllel plates of large area $A$ in ordinary physical space:
\be
E_C=-\frac{\pi^2 \hbar c}{720 \, d^3}\;A \label{casimir}
\ee
The correspondence between quantum statistical systems in physical space $(x_1,x_2,x_3)$, and classical statistical systems in 4 euclidean dimensions $(x_1,x_2,x_3,x_4)$, implies that $E_C$ can be identified with the zero-temperature limit of the classical Casimir energy between two parallel  (hyper)-planes in euclidean space at distance $d$ from each other,  having an area $A$ in, say, the $(x_1,x_2)$ plane, and an extension $\ell=\hbar c/(k_B T)$ in the fourth direction $x^4$, namely:
\be
E_C =\lim_{T \rightarrow 0}  {\cal F}^{(4)}_{pp}(d) \,\ell \,A  \;.
\ee 
Comparison with Eq. (\ref{casimir}) gives:
\be
{\cal F}^{(4)}_{pp}(d)=-k_B T \,\frac{\pi^2}{720 \, d^3}\;.
\ee
Having determined the (unit-volume) free-energy for two parallel hyper-planes in $E^{(4)}$, we can go about computing the sphere-plate energy using the PFA. To be definite, let us fix the hyper-plane to have Equation  $x_1=0$, and the three-sphere of radius $R$ to have its center at the point $x_1=d+R$ along the $x_1$ axis. Locally, the profile of the three-sphere around its tip  is described  by the height function $x_1=H(x_2,x_3,x_4)=d+(x_2^2+x_3^2+x_4^2)/2 R$. According to the PFA, in the small-distance limit $d/R \rightarrow 0$ the sphere-plate Casimir energy is estimated by taking the  average of ${\cal F}^{(4)}_{pp}(d)$ over the local separation $h$:
\be
{\cal F}_{\rm PFA}=\int dx_2 dx_3 dx_4\,{\cal F}^{(4)}_{pp}(H(x_2,x_3,x_4))= 4 \pi \int_0^{\infty} d r \,r^2 {\cal F}^{(4)}_{pp}(d+r^2/2R)\;.
\ee
Evaluation of the integral to leading order in $x$ gives the result:
\be
 {\cal F}_{\rm PFA}=-k_B T \frac{\pi^4 \sqrt{2}}{1440\,x^{3/2} } \;.\label{PFA}
\ee
We see that the PFA correctly reproduces the leading term of Eq. (\ref{EMexp}). As we pointed out in the introduction, the PFA is still widely used today to  interpret current precision small-distance Casimir experiments. It is therefore of interest to use our exact solution to check quantitatively its accuracy.  For this purpose, in Fig. \ref{figPFA} we plot the sphere-plate Casimir energy ${\cal F}/{\cal F}_{PFA}$ normalized to the PFA energy, versus $x=d/R$. It is apparent from the Figure that PFA becomes increasingly accurate at $d/R$ approaches zero. A precise perception of the accuracy of the PFA can be gained from Fig. \ref{error} where we plot (solid line) the percent error $100 ({\cal F}-{\cal F}_{\rm PFA})/|{\cal F}_{\rm PFA}|$ caused by the PFA, versus $-{\rm Log}_{10}(d/R)$.   

\section{The derivative expansion}
\label{sec::DExpa}

In recent years a new and powerful method has been put forward \cite{fosco1,bimonte1} to compute the first  curvature correction beyond PFA, i.e. the NTLO term of  the Casimir interaction between two gently curved surfaces.   In fact, the method has a general applicability, and can be used to study any short-range interaction between two surfaces.  The idea is simple to explain. In the context of  the present paper, it can be described as follows.  Consider the Casimir energy ${\cal F}$ of the system consisting of a three-plane $\Sigma$ of Equation $x_1=0$, opposed a surface described by a single-valued smooth height profile $x_1=H(x_2,x_3,x_4)$ (in general, the method can be applied also to two curved surfaces \cite{bimonte1}, but for simplicity we do not consider this more general case here).  The energy ${\cal F}$ is a functional ${\cal F}[H]$ of the height $H$. One postulates that for a small-slope surface, satisfying the condition $|\nabla H| \ll 1$ the energy functional  admits, at least upto some order in $\nabla H$, a {\it derivative expansion} (DE) in powers  of an increasing number of derivatives of $H$. To second order in $\nabla H$ one writes:
\be
{\cal F}=- k_B T \,\frac{\pi^2}{720}  \int_{\Sigma}  \frac{d^3 x} {H^3} \left[1+ \beta(H) (\nabla H)^2 +   \dots\right]\;,\label{derexp}
\ee 
where $\beta (h)$ is a pure number, and dots denote higher derivative terms. As we see, the leading term of the DE coincides with the PFA Eq. (\ref{PFA}). The coefficient $\beta$ depends in general on the  chosen field-theory, and on the bc imposed on the surfaces. We can determine the exact functional dependence of $\beta (H)$ on $H$ by comparing the gradient expansion, Eq. (\ref{derexp}), to a perturbative expansion of the Casimir energy around two flat plates, to second order in the deformation. For this purpose,  we decompose the height of the curved surface as $H(x_2,x_3,x_4)=d+h(x_2,x_3,x_4)$, where $d$ is chosen to be the distance of closest separation. For small deformations $|h(x_2,x_3,x_4)|/d \ll 1$ we can expand ${\cal F}[d+h]$ as:
\be
{\cal F}= V {\cal F}^{(4)}_{pp}(d) + \mu(d) {\tilde h}({\bf 0})+\int \frac{d^3 {\bf k}}{(2 \pi)^3} G({ k};d) |{\tilde h}({\bf k})|^2\;,\label{pert2}
\ee
where $V$ is the three-volume of $\Sigma$, ${\bf k}$ is the in-plane wave vector and ${\tilde h}({\bf k})$ is the Fourier transform of $h(x_2,x_3,x_4)$. The kernel $G(k; d)$ for the $T=0$ quantum theory in three dimensions  has been evaluated by several authors, for example in ref. \cite{goles} for a scalar field fulfilling D or N bc on both plates, as well for the EM field satisfying ideal metal bc on both plates \cite{emig3}. More recently, $G(k; d)$ was evaluated in ref. \cite{neto} for the EM field with dielectric bc. The corresponding kernel for  our classical four-dimensional euclidean theory can be simply obtained by dividing the kernel of \cite{emig3} by $\ell=\hbar c /k_B T$, and then replacing the two-dimensional in-plane vector ${\bf k} \equiv (k_1,k_2)$ of the three-dimensional theory, by the three-dimensional in-plane vectors ${\bf k} \equiv (k_1,k_2,k_3)$ of the four-dimensional theory.  For a deformation with small slope, the Fourier transform  is peaked around zero. Assuming that   the kernel can be expanded at least through order $k^2$ about $k=0$ [22], we define
\be
G(k;d)=\gamma(d)+ \delta(d)  k^2+\dots
\ee
For small $h$, the coefficients in the  DE can be matched with the perturbative result. By expanding Eq. (\ref{derexp}) in powers of $h(x_1,x_2,x_3)$ and then comparing with the perturbative expansion Eq. (\ref{pert2})  to second order in both $h$  and $k^2$, we obtain
\be
{\cal F}^{(4)'}_{pp}(d)=\mu(d)\;,\;\;\;{\cal F}^{(4)''}_{pp}(d)=2 \gamma(d)\;,\;\;\;\;\beta(d)=\frac{\delta(d)}{{\cal F}^{(4)}_{pp}(d)}\,,\label{DEpert}
\ee
where prime denotes a derivative. By using the above relations the coefficient $\beta$ for perfect conductor bc was computed in \cite{bimonte1}
\be
\beta=\frac{2}{3}\left(1-\frac{15}{\pi^2} \right)\;.\label{beta}
\ee
The values of $\beta$ for other field theories, like a scalar field satisfying D, N and mixed ND bc can be found in \cite{fosco1,bimonte1}, while for the experimentally important case of dielectric surfaces at finite temperaure $\beta$ was computed in \cite{bimonte2}.
We remark that the value of $\beta$ in Eq. (\ref{beta})  was obtained in \cite{bimonte1} by considering  the zero-temperature quantum Casimir interaction in physical space. The general equivalence between quantum statistical systems in physical space and classical statistical systems in four euclidean dimensions, implies identity of the respective $\beta$ coefficients (for same field theory and bc).  

Having determined the value of $\beta$, we can now use  Eq. (\ref{derexp}) to compute the leading correction beyond PFA to the Casimir energy between a three-plane and a three-sphere.  When doing that, one has to bear in mind that from the PFA one knows that the points of the sphere that contribute most to the Casimir interaction are those contained in a disk of radius $\sigma \simeq \sqrt{R d}$ around the sphere tip.  For small separations, it is then legitimate to  take  the power expansion of the height profile of the sphere $x_1=H(x_2,x_3,x_4)=d+(x_2^2+x_3^2+x_4^2)/2 R+ (x_2^2+x_3^2+x_4^2)^2 /8 R^3$+\dots. When this  is substituted into Eq. (\ref{derexp}), one finds 
\be
{\cal F}=-k_B T \frac{\sqrt{2}\pi^4}{1440\,x^{3/2}}   \left[ 1+\left(6 \beta -\frac{15}{4} \right)  x  +\dots \right]\;.\label{DE2}
\ee
Using the value of $\beta$ in Eq. (\ref{beta}),   we see that the DE provides the correct value for the leading correction  beyond PFA to the Casimir energy. Inclusion of this $O(x)$ correction leads to a significant improvement in the accuracy of the Casimir energy. This can be fully appreciated by looking at the dashed curve in Fig. \ref{error} which shows the corresponding percent error $100 ({\cal F}-{\cal F}_{\rm DE})/|{\cal F}_{\rm DE}|$ on the energy. Comparison with the solid line, which corresponds the PFA, reveals the significant superiority of the DE in the range $d/R < 0.01$. For example, for $d/R=0.002$, the error made by the PFA is of 0.97 \%, while the error made by the DE is -0.2 \%.

\section{The next-to-next-to-leading-order  term}  
\label{sec:DENN}

We have seen in the previous Section that the DE provides the correct value for  the first correction to the Casimir energy beyond PFA, which represents the NTLO term in the small distance expansion of the exact Casimir energy. It is interesting now to consider  NNTLO term. From Eq. (\ref{EMexp}) we see that  this term is proportional to  $\log(2 x/\pi^2)$. Compared to PFA, this term represents a  correction of order $x^{3/2} \log(x)$, which is clearly non-analytic. One may wonder if this correction  can be  computed using a higher order of the DE. The answer is no, because the DE for the em Casimir energy breaks down beyond second order in $\nabla H$. Let us assume temporarily that the DE exists beyond second order.  It is easy to convince oneself that  at the next order in $\nabla H$ the DE  involves four derivatives of the height profile.   Upto total derivatives, the most general rotationally invariant expression  involving  four derivatives of the height profile, can be recast in the  following form \cite{mazzitelli}:
$$
{\cal F}=-k_B T\frac{\pi^2}{720} \int_{\Sigma}  \frac{d^3 x}{H^3} \left[1+ \beta(\nabla H)^2 + \beta^{(1)} H^2\,(\Delta H)^2 + \beta^{(2)} H^2 \partial_i \partial_j H \partial_i \partial_j H  \right.
$$
\be
\left.+\beta^{(3)} H \Delta H (\nabla H)^2+\beta^{(4)}  (\nabla H \cdot \nabla H)^2 + \dots\right]\;,\label{DEfour}
\ee 
where dots denote again higher order terms. If Eq. (\ref{DEfour}) is used to estimate the Casimir energy of a three-sphere opposed a three-plane, following the same steps that led to Eq. (\ref{DE2}),
the result is found:
$$
{\cal F}=- \frac{k_B T\sqrt{2}\pi^4}{1440\,x^{3/2}}   \left[ 1+\left(6 \beta -\frac{15}{4} \right)  x     \right.
$$
\be
\left. -15\left(\frac{7}{32}+\frac{1}{2}\beta+\frac{144}{5} \beta^{(1)}+\frac{48}{5} \beta^{(2)}+ \frac{48}{5} \beta^{(3)}+4 \beta^{(4)} \right)  x^2+ \dots \right]\;.\label{DE4}
\ee
We see that at NNTLO the DE predicts an analytic correction to PFA of order $x^2$, instead of the correct result of order $x^{3/2} \log(x)$. The reason for the disagreement is that at fourth order the non-locality properties of the em Casimir interaction invalidate the DE expansion. This can be  proven as follows. When the fourth-order DE in Eq. (\ref{DEfour}) is matched, in their common region of validity, with the fourth-order perturbative expansion of the Casimir energy ${\cal F}[d+h]$, one finds that the coefficients $\beta^{(i)}\;,i=1,2,3,4$ occurring in Eq. (\ref{DEfour})  are  linear combinations of the coefficients of the fourth-order Taylor expansion of the perturbative Green functions $G^{(n)}({\bf k}_1,\dots,{\bf k}_n;d)$ for $n \le 4$,  in powers of the in plane momenta ${\bf k}_1,\dots,{\bf k}_n$,   about the point ${\bf k}_1=\dots={\bf k}_n={\bf 0}$. This implies that a {\it necessary} condition for the existence of the fourth order DE Eq. (\ref{DE4}) is existence of the fourth order Taylor expansion of the perturbative Green functions about zero in-plane momenta. Unfortunately, in the em case the Green functions do not admit a Taylor expansion to fourth order. Consider for example 
the second-order kernel $G(k;d)$. This kernel was computed in \cite{emig3} for the $T=0$ quantum theory in three dimensions.  As said earlier, the corresponding kernel for our classical four-dimensional euclidean theory is obtained by dividing the 3-d quantum kernel of \cite{emig3} by $\ell=\hbar c /k_B T$, and then replacing the two-dimensional in-plane vector of the three-dimensional theory by the three-dimensional in-plane vectors of the four-dimensional theory. Following \cite{emig3} the kernel is decomposed as:   
\be
G(k;d)=-\frac{2 k_B T}{d^5} [G_{\rm TM}(d k/ 2 \pi)+ G_{TE}(d k / 2 \pi)]\;,
\ee 
where $G_{\rm TM}$ and $G_{\rm TE}$ represent, respectively, the contribtutions of the transverse magnetic (TM) and transverse electric (TE) fields.
For small momenta the kernels $G_{\rm TE}(x)$ and $G_{\rm TM}(x)$ were shown in \cite{emig3} to have the power expansion:
\be
G_{\rm TM}(x)=\frac{\pi^2}{480}+\frac{\pi^4 }{1080} x^2-\frac{45+\pi^4}{6750} \pi^2 x^4 + \dots \label{GTM}
\ee
\be
G_{\rm TE}(x)=\frac{\pi^2}{480}+\frac{\pi^2(\pi^2-30)  }{1080} x^2 +\frac{\pi^3 }{32}\, x^3 -\frac{1095+50 \pi^2 +\pi^4}{6750} \pi^2 x^4 + \dots\;.\label{GTE}
\ee
Using the above expansions, and recalling Eqs. (\ref{DEpert}) it is  easily  possible to obtain the value of $\beta$ quoted in Eq. (\ref{beta}). 
The important thing to observe though is that the expansion of $G_{\rm TE}(x)$ has a contribution proportional to $x^3$, i.e. the third term on the r.h.s. of Eq. (\ref{GTE}).  The presence of this $O(k^3)$ term, which is not an analytic  function of the components $(k_1, k_2, k_3)$ of the in-plane momentum ${\bf k}$, implies that $G(k;d)$  can be Taylor expanded  about ${\bf k}=0$ only up to order $k^2$, but not to higher orders. This is just enough to ensure existence of the DE to second order in $\nabla h$, as in Eq. (\ref{derexp}),   but it invalidates the fourth-order DE in Eq. (\ref{DEfour}). The conclusion of these considerations  is that the DE can be only used to estimate the NTLO term in the Casimir energy, but it cannot be used to compute its NNTLO term. As a general remark, one may observe that the maximum possible order of the Taylor expansion of the perturbative Green functions, in powers of the in plane-momenta, ultimately  depends on how fast the perturbative Green functions fall-off to zero in coordinate space, as their arguments are taken far apart. From this perspective, is clear that existence or non-existence of the DE is a question of how  local in space the Casimir interaction is.    

As a concluding remark we observe that the fourth-order DE may exist for other more local field theories. An example is a D free scalar field. For D bc, the theory   is conformally invariant in any number $N$ of euclidean dimensions. As we said earlier, conformal symmetry was indeed exploited in \cite{eisen} to compute the  Casimir energy for a D scalar between two  (N-1)-spheres. In particular, for $N=4$ 
one finds:
\be
{\cal F}_{\rm D}=\frac{k_B T}{2}\sum_{n \ge 0} n^2 \log (1-\rho^{2 n})\;,\label{scene}
\ee 
where $\rho$ is the same quantity as in Eq. (\ref{EMene}).  By using the Plana formula, one obtains the following small distance expansion of ${\cal F}_D$ in the sphere-plate geometry:
\be
{\cal F}_{\rm D}= -\frac{\sqrt{2}\,\pi^4}{2880\,x^{3/2}} \left[1+\frac{x}{4}+\left(\frac{12}{\pi^4}-\frac{7}{480}
\right) x^2  +\left(\frac{457}{120960}-\frac{1}{\pi^4} \right)x^3+\dots\right]+ \frac{\zeta(3)}{8 \pi^2}\;.\label{scaD}
\ee
Neglecting the constant term proportional to $\zeta(3)$, which has no influence on the force, we see that the NNTLO term (i.e. the third term between the square brackets) represents an $O(x^2)$ correction, as compared the PFA. Such an order $O(x^2)$ correction is precisely what one would expect on the basis of the fourth order DE (see Eq. (\ref{DE4})).  A further argument in favour of the existence of the fourth-order DE for a D scalar comes from consideration of its  second order kernel $G_{\rm D}(k;d)$. It turns out that $G_{\rm D}(k;d)$ coincides with the em kernel for TM polarization $G_{\rm TM}$. From Eq. (\ref{GTM}) we see that, differently from $G_{\rm TE}$, $G_{\rm TM}$ does admit a fourth-order Taylor expansion in powers of the momentum about ${\bf k}=0$. Of course, to have the conclusive proof that a D-scalar admits a fourth-order DE, one would have to check the fourth-order Taylor expansions of the higher Green functions $G^{(3)}({\bf k}_1,{\bf k_2},{\bf k}_3;d)$ and $G^{(4)}({\bf k}_1,{\bf k_2},{\bf k}_3,{\bf k}_4;d)$ that are needed to determine $\beta^{(3)}$ and $\beta^{(4)}$, something that we shall not attempt here.  Finally, we note that similar to the em problem, the DE for a N free scalar breaks down at second order. The exact Casimir energy for a N scalar, which is not conformally invariant, is not known so far neither for  a system of two spheres nor for a sphere and a plate, in any number of dimensions. However, one knows that  for a N scalar the second-order kernel $G_{\rm N}(k;d)$ coincides with $G_{\rm TE}$. Thus, the same considerations done for the em problem extend to the N scalar, and show that the DE for a N scalar stops at  second order.

 \section{Conclusions}
\label{sec:conc}

We have derived a simple exact formula for the classical Casimir interaction of two perfectly conducting three-spheres in four euclidean dimensions.  Our solution includes as a special case the geometry of a three-sphere opposed a three-plane. The construction exploits the well-known conformal symmetry of vacuum Maxwell Equations in four dimensions,  together with  the conformal invariance of perfect conductor bc. Conformal invariance is used to map the system to a highly symmetric configuration of two concentric spheres, whose Casimir energy can be easily computed by the scattering formula.    The solution presented in this paper represents the only known example of an exact Casimir energy for the  full fledged Maxwell  field in a non-planar geometry. We computed the small distance expansion of the exact Casimir energy, and checked that its leading term agrees with the commonly used PFA. We verified that its NTLO term agrees with a recently proposed derivative expansion of the Casimir energy functional,  in powers of an increasing number of derivatives of the height profile of the sphere. We  find that the NNTLO  represents a correction of  order $(d/R)^{3/2} \log(d/R)$ compared to the PFA, where $d$ is the separation and $R$ the sphere radius. By examining the singularities of the perturbative kernel of the Casimir energy, to second order in the height profile,   we prove that the derivative expansion breaks down beyond second order, and therefore it cannot be used to compute the NNTLO correction.   

\appendix
\label{sec:appen}

\section{Scattering by a perfectly conducting three-sphere}

 In this Appendix we briefly discuss scattering of em waves by a perfectly conducting three-sphere $S^{(3)}$ of radius $R$, in four euclidean dimensions. Let $\phi_{\rm id}: S^{(3)} \rightarrow E^{(4)}$ the identity map on $S^{(3)}$ providing the embedding of the three-sphere into $E^{(4)}$.  Perfect-conductor bc on $S^{(3)}$ can be conveniently expressed by the condition
\be
(\phi_{\rm id}^*F)_{ab}=0 \;,\label{bc2}
\ee  
where $\phi^*$ denotes the pull-back of the map $\phi$, and  $F_{ab}$ is the Maxwell field strenght (we follow the abstract index notation of tensors \cite{wald}. According to this notation latin indices from the initial part of the alphabet a,b,c,... shall denote abstract tensor indices, while greek letters $\mu, \nu...$ shall label coordinate components of tensors. Letters form the middle latin alphabet i,j,... shall label angular coordinates on the unit three-sphere).  The above bc  constitute the natural generalization of the familar bc for perfect conductors, requiring vanishing of the tangential component of the electric field and of the normal component of the magnetic field on the surface of a perfect conductor. 

Spherical symmetry allows to easily solve the scattering problem. For this purpose, it is useful to use a spherical vector wave basis.  To construct it, we consider in the euclidean space $E^{(4)}$ a spherical coordinate system $\{x^{\mu}\}\equiv\{ \alpha^1,\alpha^2,\alpha^3,r\}$ with origin  at the center of the  three-sphere. Here, $r$ is the radial coordinate, and $\alpha^1,\alpha^2,\alpha^3$ are angular coordinates on the unit three-sphere, such that the euclidean line-element takes the form:
\be
ds^2=dr^2+r^2 d \Omega^2_3\;,
\ee 
where
\be
d \Omega^2_3=\Omega_{ij} d\alpha^i\,d\alpha^j
\ee
is the metric on the unit three-sphere. It is convenient to express the field strength $F_{ab}$ in terms of the potential $A_{a}$:
\be
F_{ab}=\partial_{a}\,A_{b}- \partial_{b}A_{a}\;.
\ee 
The potential $A_{a}$ can be chosen to satisfy the following  transversality gauge condition:
\be
A_r=0\;,\;\;\;\;\;\nabla_i \,A^i=0\;.
\ee  
The transverse potential $A_i$ admits a generalized Fourier expansion in terms of an orthonormal basis of vector hyperspherical harmonics $S_{i|nlmp}(\alpha^1,\alpha^2,\alpha^3)$, which are eigenfunctions of the vector Laplacian on the unit three-sphere \cite{louko}: 
\be
A_i=\sum_{nlmp} f_{nlmp}(r) \, S_{i | nlmp}(\alpha^1,\alpha^2,\alpha^3)\;,\label{four}
\ee
with coefficients $f_{nlmp}(r)$  that depend only on the radial coordinate $r$. The index $n$ takes values $2,3,\dots$, and the overall degeneracy described by the indices $l,m$ and the parity index $p$ is $2(n^2-1)$ (for details see \cite{louko}). Plugging the expansion (\ref{four}) into Maxwell Equations, one finds that the radial dependence of the coefficients $f_{nlmp}$ is of the form:
\be
f_{nlmp}= \frac{a}{r^n}+ b \,r^n\;,\label{radial}
\ee 
where $a$ and $b$ are arbitrary constants. Using the above results, the scattering problem for a perfectly conducting three-sphere of radius $R$ is easily solved. We note first that the bc Eq. (\ref{bc2}) are equivalent to the condition:
\be
(\phi_{\rm id}^*A)_{a}=0 \;,\label{bc3}
\ee 
which simply say that the transverse vector potential  vanishes on $S^{(3)}$. We need  distinguish  two different scattering problems, i.e. external and internal scattering. In the external problem the incoming wave $A_i^{({\rm in})}$ originates outside the three-sphere and therefore it  must be regular at all points {\it inside} the three-sphere. On the contrary the scattered   wave $A_i^{({\rm scat})}$ must be regular {\it outside} the sphere and vanish at infinity.  These conditions imply that for the incoming wave the coefficient $a$ in Eq. (\ref{radial}) must vanish, while for the scattered wave the coefficient $b$ must be zero.  We shall normalize the incoming waves such that $b^{({\rm in})}=R^{-n}$, and $a^{({\rm scat})}=R^{n}$:
\be
A_{i|nlmp }^{({\rm in|ext})}=\left(\frac{r}{R} \right)^n \, S_{i | nlmp}(\alpha^1,\alpha^2,\alpha^3) \;,\;\;\;\;A_{i|nlmp }^{({\rm scat|ext})}=\left(\frac{R}{r} \right)^n \, S_{i | nlmp}(\alpha^1,\alpha^2,\alpha^3)\;.\label{exter}
\ee
After we impose the bc Eq. (\ref{bc3}) on the total field $A_i=A_i^{({\rm in})}+A_i^{({\rm scat})}$,  it is easy to verify that the scattering amplitude for the external problem  is diagonal with elements: 
\be
{\cal T}_{nlmp,n'l'm'p'}=-\delta_{nn'} \delta_{l l'} \delta_{mm'} \delta_{pp'}\;.\label{scatmat}
\ee 
Consider now the {\it internal} scattering problem. Now the incoming wave $A_i^{({\rm in})}$ originates    from a point inside the three-sphere, and therefore it must vanish at infinity, while
the scattered   wave $A_i^{({\rm scat})}$ must  be regular at all points {\it inside} the three-sphere. These requirements imply that for the incoming wave the coefficient $b$ in Eq. (\ref{radial}) must vanish, while for the scattered wave the coefficient $a$ must be zero.  We shall normalize the incoming waves such that $a^{({\rm in})}=R^{n}$, and $b^{({\rm scat})}=R^{-n}$: 
\be
A_{i|nlmp }^{({\rm in|int})}=\left(\frac{R}{r} \right)^n \, S_{i | nlmp}(\alpha^1,\alpha^2,\alpha^3) \;,\;\;\;\;A_{i|nlmp }^{({\rm scat|int})}=\left(\frac{r}{R} \right)^n \, S_{i | nlmp}(\alpha^1,\alpha^2,\alpha^3)\;.\label{inter}
\ee
It easily follows from the above Equation that  the internal scattering amplitude coincides with the external one in  Eq. (\ref{scatmat}).

\acknowledgments

The author thanks T. Emig, N. Graham, M. Kruger, R. L. Jaffe and M. Kardar for valuable discussions while the manuscript was in preparation.



\end{document}